\documentclass{nature}
\usepackage{etoolbox}
\usepackage{bm}
\usepackage{amssymb}
\usepackage{epsfig}
\usepackage{color}
\usepackage{amssymb,amsmath}
\usepackage{aas_macros}
\usepackage{ulem}
\usepackage{graphicx,epstopdf}
\usepackage{url}

\title{No hot and luminous progenitor for Tycho's supernova}
\author{ T.~E. Woods$^{1*}$, P. Ghavamian$^{2}$, C. Badenes$^{3}$, M. Gilfanov$^{4,5}$}
\date{}
\begin{document}
\maketitle
\pagenumbering{gobble}
\begin{affiliations}
\item Monash Centre for Astrophysics, School of Physics and Astronomy, Monash 
University, VIC 3800, Australia
\item Department  of  Physics,  Astronomy  and  Geosciences,  Towson University, Towson, MD, 21252, USA
\item Department of Physics and Astronomy and Pittsburgh Particle Physics, Astrophysics, and Cosmology Center, University of Pittsburgh, 3941 O'Hara Street, Pittsburgh, PA, 15260, USA
\item Max-Planck Institut f{\"u}r Astrophysik, Karl-Schwarzschild-Str. 1, D-85741 Garching, Germany
\item Space Research Institute, Profsoyuznaya 84/32, 117997 Moscow, Russia
\end{affiliations}

\begin{abstract}
Type Ia supernovae have proven vital to our understanding of cosmology, both as standard candles and for their role in galactic chemical evolution; however, their origin remains uncertain. The canonical accretion model implies a hot and luminous progenitor which would ionize the surrounding gas out to a radius of $\sim$10--100 parsecs for $\sim$100,000 years after the explosion. Here we report stringent upper limits on the temperature and luminosity of the progenitor of Tycho's supernova (SN 1572), determined using the remnant itself as a probe of its environment. Hot, luminous progenitors that would have produced a greater hydrogen ionization fraction than that measured at the radius of the present remnant ($\sim$3 parsecs) can thus be excluded. This conclusively rules out steadily nuclear-burning white dwarfs (supersoft X-ray sources), as well as disk emission from a Chandrasekhar-mass white dwarf accreting $\gtrsim 10^{-8}M_{\odot}$yr$^{-1}$ (recurrent novae).  The lack of a surrounding Str{\" o}mgren sphere is consistent with the merger of a double white dwarf binary, although other more exotic scenarios may be possible.
\end{abstract}

Four hundred and forty-five years ago, the explosion of SN 1572 in the constellation of Cassiopeia demonstrated definitively that the night sky was not permanent -- the Universe evolves. The nature of this supernova, however, was not conclusively determined until recently, when modern analysis of the historical light curve,\cite{Ruiz_Lapuente04} the X-ray spectrum of the supernova remnant\cite{Badenes06} and the spectrum of light echoes from the explosion scattered off of interstellar dust\cite{Krause08} showed that it belonged to the majority class of Type Ia supernovae (SNe Ia). In spite of these advances, however, the nature of the star which gave rise to this explosion, as well as all others like it, remains unknown. Theoretical models fall into two broad categories: the accretion scenario,\cite{WI73} wherein a white dwarf grows slowly in mass through accretion and nuclear-burning of material from a binary companion prior to explosion; and alternatively, the merger scenario,\cite{Webbink84} wherein a binary pair of white dwarfs merge after shedding angular momentum through gravitational-wave radiation. Efforts to constrain the nature of the progenitor and any surviving companion star in the vicinity of Tycho's supernova\cite{RuizLapuente04,Zhou16} have remained inconclusive.\cite{Kerzendorf09,MMN14} In particular, there remains significant disagreement on the location of the centre of Tycho's supernova, and therefore the viability of any candidate surviving donor star.\cite{Bedin14,Williams16} The extent to which the most commonly cited candidate, Tycho g, stands out from other stars along the same line of sight also remains in question.\cite{GH09,Kerzendorf13}

Attempts to detect emission from an individual SN Ia progenitor system have thus far relied largely on searching pre-explosion images of the host galaxies of a few nearby and very recent events.\cite{Nielsen12,Graham15} These searches have provided some constraints on the presence of hot luminous progenitors immediately prior to explosion, as well as optically-luminous companions, for these few SNe Ia -- in varying degrees of tension with the accretion scenario. Using observations from the Kepler spacecraft, red giant donors were also ruled out for three likely type Ia supernovae in red, passive galaxies,\cite{Olling15} based on the lack of observed shock interaction between the expanding supernova and a Roche-lobe filling companion. However, 
SNe Ia in passive galaxies are not representative of the ``typical'' population.\cite{MMN14} None of these observations can make any statement about the earlier evolution of the progenitor. Nor can these constraints be applied to the large sample of resolved supernova remnants in the Local Group. Conversely, searches for evidence of any luminous progenitor population in X-ray\cite{GB10,DiStefano10} or UV\cite{WG13,Johansson16} emission have placed strong constraints on the total contribution of such progenitors to the observed SN Ia rate in nearby galaxies, but this cannot be inverted to yield information on the origin of individual objects.

Here we propose an alternative test: to search for a ``fossil'' or ``relic'' nebula around individual supernovae photoionized by their progenitors. In the accretion scenario, the process which leads to the growth of the white dwarf mass is fusion of hydrogen to helium and further to carbon and oxygen. These accreting white dwarfs must go through a long-lived ($\gtrsim$ 100,000 years), hot luminous phase of steady nuclear burning on the surface, with effective temperatures of $10^{5}$--$10^{6}$K and luminosities of $10^{37}$--$10^{38}$erg/s at some point prior to explosion.\cite{Rappaport94} Such objects are expected to be significant sources of ionizing radiation.\cite{Rappaport94,WG16} 

For accretion rates $\lesssim \rm{few} \times 10^{-7}M_{\odot}$ yr$^{-1}$, hydrogen fusion is subject to thermal instability and the nuclear energy is mostly released in the 
form of classical and recurrent nova explosions.\cite{Prialnik95,Nomoto07,Wolf13} Recent measurements of the WD mass in a few nearby recurrent novae yielded values close to the Chandrasekhar mass limit,\cite{Thoroughgood01,Darnley15} suggesting that these objects may contribute to the production of Type Ia supernovae (if they are carbon-oxygen white dwarfs). However, {\bf any} accretion onto a white dwarf must also release gravitational potential energy. In the parameter range expected for Type Ia supernova progenitors, accretion proceeds through an optically-thick, geometrically-thin (Shakura-Sunyaev)\cite{SS73} disk, which has a well-understood
luminosity, temperature, and emitted spectrum (see Methods). From this, one finds that for massive white dwarfs undergoing accretion rates as low as $\sim10^{-8} M_{\odot}$ yr$^{-1}$, the typical inner disk temperatures and 
luminosities are on the order of $10^{5}$K and $10^{35}$erg/s, rendering them much dimmer (though not insignificant) sources of ionizing radiation.

Any nebula created by the supernova progenitor will persist until sufficient time has passed for the majority of the gas to recombine.\cite{BKS70} This can be estimated from the hydrogen recombination rate $\alpha_{B} (H^{0},T)$, and the density $n_{\rm ISM}$ of the surrounding interstellar medium, assuming the gas is initially nearly wholly ionized (such that the electron density $n_{\rm e} \sim n_{\rm ISM}$):

\begin{equation}
\tau_{rec} = \left(n_{\rm{e}} \alpha_{B} (H^{0},T\,\approx\,10^{4}K)\right)^{-1} \approx (100,000) \times \left(\frac{n_{\rm{ISM}}}{1\,\rm{cm}^{-3}}\right)^{-1}\,\,\,\rm{years}
\end{equation}

\noindent where we have assumed Case B recombination (i.e., ionizing photons produced by recombinations to the ground state are immediately absorbed). If the medium is initially only partially-ionized, it will have a longer characteristic recombination timescale. Typically $n_{ISM}\,\sim\,$1 cm$^{-3}$,  so $\tau_{rec}\,\sim$ 100,000 years. This allows one to constrain the ionization history of supernova progenitors by searching for evidence of their lingering impact on their environment, thus enabling “Type Ia supernova archaeology.”\cite{WG16} This requires knowledge of the density and ionization state of the ISM in the $\sim$ 1 -- 100 pc vicinity of known SNe Ia.\cite{Graur14}

One way to obtain this information is to use the expanding supernova shock itself as a probe of the surrounding ISM.  Tycho's supernova remnant (SNR) is one of a number of known SNe Ia remnants whose forward shock is traced in part by filaments of Balmer line optical emission. This arises due to collisional excitation of neutral hydrogen immediately behind the advancing shock, where excitation of cold neutral hydrogen produces a narrow Balmer emission line, while excitation of hot neutral hydrogen formed by charge exchange gives rise to a broad Balmer emission line.\cite{CKR80} {\it The very existence of this emission along the eastern and northern periphery of Tycho's SNR demonstrates that the ambient environment around the remnant (and thus by extension its progenitor) is at least partially neutral.}\cite{CKR80, Ghavamian00} Modelling of both the H$\alpha$ and H$\beta$ broad-to-narrow flux ratios from the forward shock as well as the [O~III]/H$\beta$ ratio from the photoionization precursor ahead of the forward shock indicates that the ambient hydrogen must be at least 80\% neutral in this region\cite{Ghavamian00,Ghavamian01} (see also Supplementary Information).  This strongly constrains the ionizing luminosity from the progenitor prior to, and during, the explosion.\cite{CR78,Ghavamian03,Vink12}

Recently, it has been suggested based on CO observations\cite{Zhou16} that Tycho's SNR is associated with dense clumps of molecular gas in the same region of the Milky Way, suggesting a thick molecular shell possibly excavated by a fast, continued outflow from the progenitor.
We note, however, that the known age (445 yr), physical size ($\sim$3 pc)\cite{Yamaguchi14} and ionization timescale (log ($\rm{n}_{\rm{e}}$t/($\rm{cm}^{-3}$s)) $\sim$10.5 for the Si ejecta)\cite{Badenes07} of Tycho confidently rule out an expansion into any kind of low-density cavity or dense, massive shell excavated by a progenitor outflow.\cite{Badenes07, PB17} 
These properties of the SNR are fully consistent with an expansion into an undisturbed ISM with average density n$_{ISM}\,\sim$0.5 to 1 $\rm{cm}^{-3}$ since the time of the explosion.  This is very close to the density that the outer shock is running into today (although the remnant is also encountering denser gas on the eastern edge, this is not characteristic of the mean density of the environment).\cite{Williams13} Any large-scale modification of the ambient medium around the SN is in direct conflict with the bulk properties of the SNR (for more details, see Figure 3 and discussion in Patnaude \& Badenes 2017). Additional arguments against the existence of any molecular bubble associated with the remnant of Tycho's supernova include the spatial extent of the observed photoionization precursor, and the marked discrepancy between the velocities of the CO and H Balmer emission lines measured relative to the local standard of rest. These arguments are summarised in the Supplementary Information. We note however, that even in the event such a molecular bubble were associated with Tycho, with an inner radius just outside the present radius of the shock, the low density and ionization state of the gas interior to this point which is presently being overrun would still provide the same constraint on the nature of the progenitor prior to explosion.

The characteristic size of the  nebula ionised by the hypothetical hot supernova progenitor is determined by the ``Str{\" o}mgren'' radius ($\rm{R}_{\rm{S}}$), which scales as\cite{WG16}:
\begin{equation}
\rm{R}_{\rm{S}} \approx 35\rm{pc}\left(\frac{\dot N_{\rm{ph}}}{10^{48}\,\rm{s}^{-1}}\right)^{\frac{1}{3}}\left(\frac{\rm{n}_{\rm{ISM}}}{1\,\rm{cm}^{-3}}\right)^{-\frac{2}{3}}\label{RS}
\end{equation}
\noindent where $\dot N_{\rm{ph}}$ is the ionizing luminosity (in photons per second). Note that for variable sources, a weighted average of the ionizing photon luminosity over the recombination timescale is the quantity of interest.\cite{CR96} The number of ionizing photons emitted per unit energy depends on the shape of the emitter's spectrum, but for photospheric temperatures in the range $2\times10^{4}$ K $\lesssim$ T $\lesssim$ $10^{6}$ K it is $\sim10^{9}$--$10^{10}$ ionizing photons/erg.

\begin{figure}
\vspace{-1cm}
\caption{Hydrogen ionization fraction as a function of radial distance from the progenitor of Tycho's supernova, for putative objects with effective temperatures of $10^{5}$K (blue) and $10^{6}$K (red). Dashed lines denote a source with a bolometric luminosity of $10^{36}$erg/s, solid lines denote a source with a bolometric luminosity of $10^{38}$erg/s. These values roughly bracket the range in expected accreting, nuclear-burning white dwarf temperatures and luminosities.\cite{Greiner00} Shown in black is the upper limit on the hydrogen ionization fraction at the present radius of the shock.}\label{Hfrac}
\begin{center}
\includegraphics[width=0.9\textwidth]{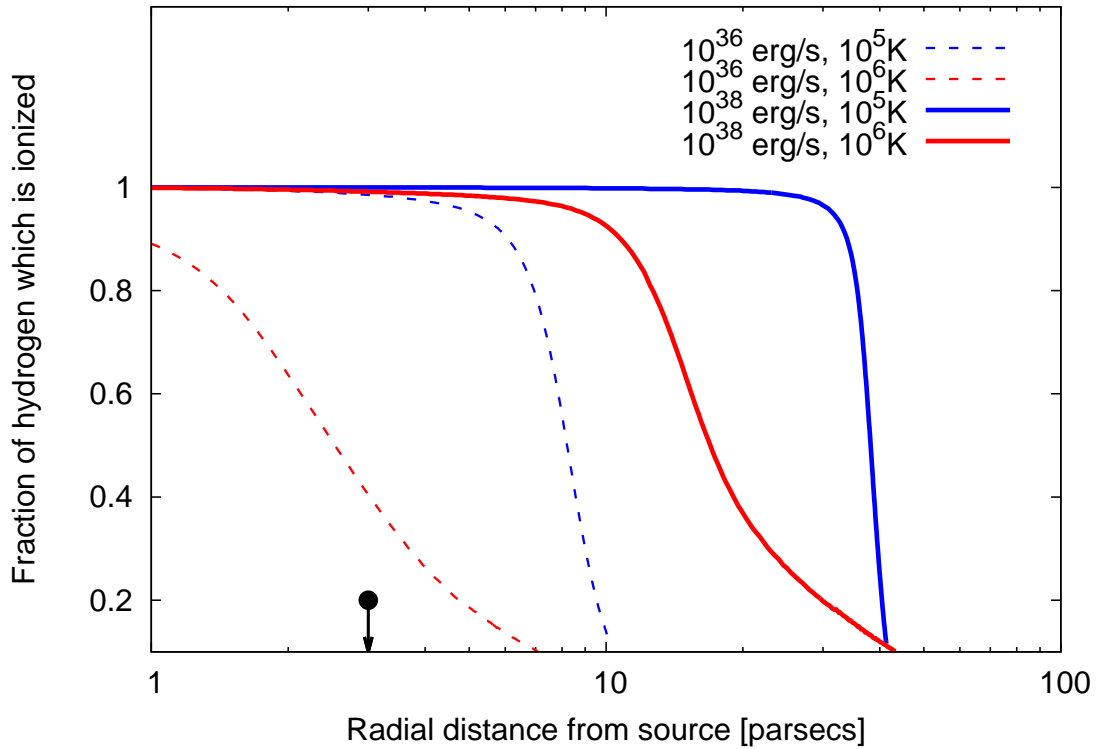}
\end{center}
\end{figure}

For relatively cooler ionizing sources (T $\lesssim10^{5}$ K), the outer boundary of the ionized nebula is very sharply defined (see blue lines in Fig. \ref{Hfrac}), owing to the high photoionization cross-section for hydrogen.  Therefore, given the ambient density of the surrounding ISM inferred above, the presence of {\bf any} neutral hydrogen at the forward shock radius of Tycho's SNR ($R_s\lesssim 3$ pc) places a strict upper limit on the size of the ionised nebula for Tycho's supernova. For an average surrounding ISM density of $n_{ISM}\,\lesssim 1 \rm{cm}^{-3}$, this translates to an upper limit on the ionising photon luminosity of $\dot N_{\rm{ph}}$ $\lesssim 6\times 10^{44}\rm{~s}^{-1}$. From eq. \ref{RS}, the upper limit on the ionising source luminosity scales as the $\propto n_{ISM}^2$. However, as explained above, the density of the ISM surrounding Tycho's supernova is fairly well constrained.

Hotter sources ($\rm{T}_{\rm{eff}}$ $\gtrsim$ few $\times 10^{5}$ K) produce higher energy photons with longer mean free paths, which broaden the boundary between ionized and neutral media.  This necessitates using a detailed photoionization simulation in order to determine the fraction of ionized hydrogen as a function of radius (see Methods). This is illustrated in Fig. \ref{Hfrac}, from which it is clear that any luminous ($\gtrsim 10^{36}$erg/s), high temperature source would still have produced a greater ionized hydrogen fraction ($\approx\,$20\%) than observed at the present radius of the remnant. Here we have assumed $n_{ISM}\,\approx\,$1 $\rm{cm}^{-3}$, approximately the observed upper limit. Lower densities would yield larger ionized regions (cf. eq. \ref{RS}).

\begin{figure}
\vspace{-1cm}
\caption{Upper limits on the typical luminosity of the progenitor of Tycho's supernova (SN 1572) during the last 100,000 years. Thick black lines denote theoretical accreting stable nuclear-burning white dwarf models.\cite{Wolf13} Thin black boxes mark the approximate ranges of temperatures and luminosities inferred for confirmed close-binary Magellanic supersoft sources with definitive luminosity measurements\cite{Greiner00,Starrfield04}: 1.~CAL 87; 2.~1E 0035.4-7230; 3.~RX J0513.9-6951; and 4.~CAL 83. Note that the only source that appears to be inconsistent with theoretical models is CAL 87; however, this source is viewed very nearly edge-on, and the disk is understood to at least partly obscure the central hot object.\cite{Ness13}}\label{mainBB}
\begin{center}
\includegraphics[width=0.9\textwidth]{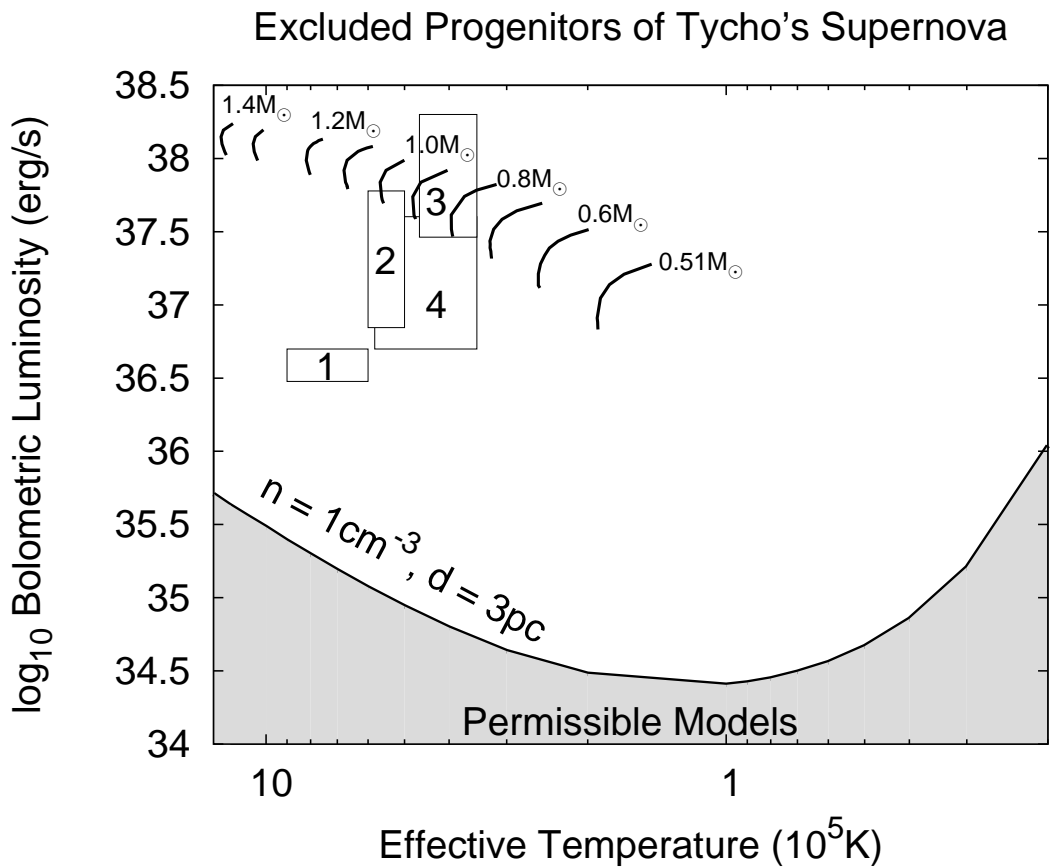}
\end{center}
\end{figure}

We summarize our constraints on hot, luminous progenitors in Fig. \ref{mainBB}, which compares our upper limits on the luminosity as a function of effective temperature with theoretical models of white dwarfs accreting at rates capable of sustaining steady nuclear-burning of hydrogen.\cite{Wolf13}  For comparison, we include parameters for several observed sources.\cite{Greiner00} All are confidently excluded. Note that for putative progenitors with complex accretion histories, any arbitrary trajectory in the HR diagram given in Fig. \ref{mainBB} can be excluded using the same constraint i.e., if it produces too great a time-averaged photoionizing flux. From these results it is clear the progenitor of Tycho's supernova cannot be described by the classic nuclear-burning accretion scenario. A white dwarf accreting at a much higher rate, such that it ejected sufficient mass in a fast wind that might have masked the ionizing flux,\cite{HKN96,NG15} would have strongly modified the surrounding environment, in conflict with the apparent evolution of the shock into an undisturbed, constant density medium.\cite{Badenes07} Slow ($\sim 100$km/s) winds could in principle obscure the lowest luminosity sources we otherwise exclude in Fig. \ref{mainBB} ($\sim 10^{36}$erg/s) for mass loss rates greater than $10^{-8} M_{\odot}$/yr (e.g., from a companion on the first giant branch)\cite{Cumming96}; however, radio\cite{PerezTorres14,Chomiuk16} and X-ray\cite{Margutti12,Margutti14} observations have ruled out such winds in the environments of normal SNe Ia,\cite{Kundu17} and no surviving giant donor consistent with this scenario has been found for Tycho (see discussion above).

We can perform a similar experiment using our photoionisation simulations, given emission spectra from an accretion disk around a white dwarf. For a Chandrasekhar-mass white dwarf, the threshold for an accretion rate capable of steady nuclear-burning is $\sim 4\times 10^{-7} ~M_\odot$ yr$^{-1}$.\cite{Nomoto07,Wolf13} We find that any Shakura-Sunyaev\cite{SS73} disk with accretion rates exceeding $\gtrsim 10^{-8}M_{\odot}$ yr$^{-1}$ onto a Chandrasekhar-mass white dwarf can be confidently excluded. For hydrogen-accreting white dwarfs, this is approximately the threshold accretion rate below which the mass ejected in novae is theoretically expected to exceed that accumulated between outbursts -- i.e., the white dwarf could not have been growing in mass.\cite{Yaron05} Note that, more generally, our constraint on the accretion rate is independent of whether the white dwarf is accreting hydrogen or helium. Thus, the detection of neutral matter in the vicinity of Tycho's supernova is strongly constraining also for accreting scenarios without surface nuclear burning. In particular, it excludes any nova progenitor with recurrence time shorter than $\sim 50$ years.\cite{Nomoto07,Wolf13}

To conclude, we rule out steadily nuclear-burning white dwarfs or recurrent novae as the progenitors of Tycho's supernova. This is consistent with recent theoretical work indicating sigificant mass accumulation in steadily hydrogen-burning accreting white dwarfs may not be feasible.\cite{Denissenkov17} Models which do not predict a hot, luminous phase prior to explosion, such as the merger or ``double-degenerate'' scenario, remain consistent with our result. This includes so-called ``violent'' mergers,\cite{Pakmor13} although such explosions may be expected to be too asymmetric to explain typical SNe Ia,\cite{Bulla16} including Tycho's supernova.\cite{Williams17} Notably, it has been suggested that some white dwarf mergers may actually produce a short-lived soft X-ray source; this too is excluded by our constraint, although the same theoretical models suggest in these instances the object may not explode as a SN Ia.\cite{Shen12,Schwab16} We also cannot exclude a `spin-up-spin down' single-degenerate progenitor model with a spin-down timescale longer than $\sim10^5$ years for the origin of Tycho's supernova,\cite{Justham11,DiStefano11,Benvenuto15} although there remain other theoretical and observational challenges for this scenario.\cite{MMN14}

Given that the light echo spectrum of Tycho's supernova has revealed it to be a typical Type Ia,\cite{Krause08} any plausible model for the origin of the majority of SNe Ia must remain consistent with the constraint outlined here. Similarly strong constraints -- or detections -- can be obtained for other nearby SN Ia remnants with sufficiently deep observations,\cite{WG16} using the expanding shock to probe the progenitor's environment. This opens a new path to reveal at last the progenitors of SNe Ia.

\section*{Bibliography}
\bibliographystyle{naturemag}

\section*{Corresponding author}

Correspondence to Tyrone E. Woods (tyrone.woods@monash.edu).

\section*{Acknowledgements}

The work of P. G. was supported by grants HST-GO-12545.08 and HST-GO-14359.011. C. B. acknowledges support from grants NASA ADAP NNX15AM03G S01 and NSF/AST-1412980. M. G. acknowledges partial support by Russian Scientific Foundation (RNF) project 14-22-00271.

\section*{Contributions}

T.~E.W. lead the Cloudy simulations and analysis of their results, and was the primary author of the main text and methods. P.G. wrote the supplementary section of the paper, and wrote portions of the main manuscript summarizing the constraints on preshock conditions from the Balmer-dominated shocks. C.B. first suggested this project during the conference ‘Supernova Remnants: An Odyssey In Space After Stellar Death’ in Crete, and contributed to the text and the interpretation of the analysis. M.G. contributed to defining the simulations setup, analysis and interpretation of Cloudy results and to the writing of the manuscript.

\section*{Methods}

\subsection{Photoionization models:}

Given that we are considering relatively high-temperature ($>10^{5}$K) ionizing sources, with correspondingly broader transition regions between ionized and neutral media than given analytically by the classical Str{\" o}mgren boundary, we model the size of any putative photoionized region using the plasma simulation and spectral synthesis code {\sc Cloudy} v13.03.\cite{Ferland13}

{\sc Cloudy} determines the gas temperature, ionization state, chemical structure, and emission spectrum of a photoionized nebula by solving the equations of statistical and thermal equilibrium in 1-D. The code relies on a number of critical databases for its calculations; notably tables of recombination coefficients,\cite{Badnell03,Badnell06} and ionic emission data taken from the CHIANTI collaboration database version 7.0.\cite{Dere97,Landi12}

We assume spherical symmetry in our models with a surrounding ISM having density $n_{ISM}\,=\,$1 $\rm{cm}^{-3}$, unless otherwise stated.  Lower densities would only result in larger nebulae for fixed source temperature and luminosity, and $n_{ISM}\,\approx 1\, \rm{cm}^{-3}$ is the approximate upper bound inferred for the pre-shock intercloud ISM in the vicinity of Tycho's SN remnant. We assume solar metallicity for the ISM in the vicinity of Tycho; the default solar values as defined in {\sc Cloudy} are taken from Grevesse \& Sauval (1998),\cite{GS98} with updates to the oxygen and carbon abundances\cite{AP01,AP02} as well as those of nitrogen, neon, magnesium, silicon, and iron.\cite{Holweger01}  Modest variations in the metallicity will not significantly effect the radius of any photoionized nebula. Given that the age of Tycho's supernova remnant is $<< \tau _{\rm{rec}}$, we assume steady-state models throughout.

We approximate the spectra of nuclear-burning white dwarfs as blackbodies. This provides a reasonable fit to their {\bf ionizing} emission,\cite{WG13,WG14,WG16} with significant deviations only arising far into the Wien tail. For white dwarf accretion disks, we use the ezDiskbb\cite{Zimmerman04} model from the X-ray spectral modelling software package {\sc xspec}\cite{XSPEC} to produce Shakura-Sunyaev\cite{SS73} disk spectra for any desired white dwarf mass and accretion rate. This sets the shape of the spectrum in {\sc cloudy}. We normalize the luminosity of the accretion disk to the rate of gravitational potential energy release:

\begin{equation}
L = \frac{1}{2}\frac{GM_{\rm{WD}}\dot M}{R_{\rm{WD}}}
\end{equation}

\noindent for a given white dwarf mass ($M_{\rm{WD}}$), radius ($R_{\rm{WD}}$), and accretion rate ($\dot M$). For the white dwarf radius, we approximate numerical results for a zero-temperature white dwarf\cite{Panei00} radius with the relation\cite{GB10}:

\begin{equation}
R_{\rm{WD}}    = 0.0126\left(\frac{M_{\rm{WD}}}{M_{\odot}}\right)^{-\frac{1}{3}}\left(1.0 - \left(\frac{M_{\rm{WD}}}{1.456M_{\odot}}\right)^{\frac{4}{3}}\right)^{\frac{1}{2}}R_{\odot}
\end{equation}

\noindent We conservatively adopt the radius of an approximately 1.35 $M_{\odot}$ carbon-oxygen white dwarf in producing our limit on accreting objects. Any more massive white dwarf would have a smaller radius and thus larger disk luminosity. 

Note that we do not include emission from the boundary layer. For slowly rotating white dwarfs, the luminosity in the boundary layer should be comparable to that of the disk. If the boundary layer is optically-thick, this should roughly double the total ionizing luminosity. This is expected on theoretical grounds.  However, given the difficulty in matching this to observed cataclysmic variables, we do not include an optically-thick boundary layer in our estimates.

\noindent {\bf Data availability}: The photoionization and spectral synthesis code Cloudy used in this work is open-source, and may be downloaded from \url{www.nublado.org}. The data that support the plots within this paper and other findings of this study are available from the corresponding author upon reasonable request.

\section*{Bibliography}
\bibliographystyle{naturemag}

\section*{Supplementary Information}
\subsection{The interstellar environment surrounding Tycho's supernova}


The optically emitting shocks in Tycho's supernova remnant are located along its eastern and northeastern edges, and appear as Balmer-dominated filaments.  At these locations, the forward shock is 
propagating into a strong density gradient with a density nearly 10 times larger than the rest of the (non-optically emitting) SNR.\cite{Williams13}  The brightest of the optical filaments 
in Tycho's SNR is known as  ``Knot g''.\cite{KvdB78}  A thick shell of diffuse optical emission is observed extending ahead of the Balmer-dominated shocks.\cite{Ghavamian00}  Models of the expansion of the remnant\cite{Badenes07} and both X-ray and infrared observations,\cite{Yamaguchi14, Williams16} combined with the mass swept up by the forward shock, indicate that for most of its 
existence Tycho's SNR must have propagated through a low density ($\lesssim 0.5 \,\rm{cm}^{-3}$) environment.  This is typical of the warm ionized/warm neutral interstellar medium.  

The diffuse emission extending ahead of the Balmer filaments in Tycho has been identified as a photoionization precursor produced by He~II 304 \AA\, photons
(He~II Lyman $\alpha$) from the Balmer-dominated shocks.\cite{Ghavamian00} The He~II emission is collisionally excited behind the Balmer-dominated shocks,
which move  at $v_{sh}\,\sim$\,2000 km/s.   At 40.8 eV per photon, the He~II 304 \AA\, radiation field from these shocks is both
energetic and dilute, causing the precursor to remain simultaneously under-ionized (as evidenced by a low [O III]/H$\beta$ ratio, $\sim$ 1)\cite{Ghavamian00} and hot (H$\alpha$ line width 
$\sim$30 km/s, measured from high resolution spectroscopy).\cite{Lee07}  Due to its low density, the precursor gas fails to achieve ionization equilibrium before being overrun by the forward shock.\cite{Ghavamian00}   Further evidence of a high neutral fraction was found by Ghavamian et al. (2001), who 
modeled the ratio of broad to narrow Balmer line emission from Knot g and found that the observed ratio of broad to narrow flux in H$\alpha$ and H$\beta$ required an initial ionized hydrogen fraction $\rm{f}_{\rm{H II}}$ $<$ 0.2 ($\rm{f}_{\rm{N}} > 0.8$).\cite{Ghavamian01}

The spatial extent of the photoionization precursor is expected to be on the order of one mean free path for photoionization 
of hydrogen by He~II $\lambda$304 \AA\, photons, or $\ell_{mfp}\,\sim\,(n_{HI}\sigma_i(304))^{-1}$.  Williams et al. (2013)\cite{Williams13} estimated the postshock density along the full circumference of Tycho's SNR by modeling its dust emission from mid-infrared imagery with the Spitzer Space Telescope.  Assuming a factor of 4 compression by the Balmer-dominated shocks, their estimates yield a preshock density in the range 1.0 - 5.0 cm$^{-3}$ for the Balmer filaments (recall that these larger values are consistent with a density gradient along the eastern side of Tycho, and are not representative of the much lower mean preshock density of 0.5-1 cm$^{-3}$ averaged along the rim).  Combining this estimate with the neutral fractions from above, this yields a spatial scale $\sim$ 0.3-1.4 pc for the photoionization precursor.  This corresponds to a size $\sim$0.3$^{\prime}$ - 1.6$^{\prime}$ for an assumed remnant distance of 3 kpc.  This is in good agreement with the observed scale length of the diffuse H$\alpha$ emission.\cite{Ghavamian01,Lee07} Together with the density constraints described from evolutionary models in the text, conditions in the interstellar medium surrounding Tycho's SNR have been well constrained. 
\subsection{Association (or lack thereof) between Tycho's supernova and molecular clouds}

In their CO line maps of the environs of Tycho's SNR, Zhou et al. (2016)\cite{Zhou16} found enhanced $^{12}$CO $J\,=\,2-1$ emission relative 
to $^{12}$CO $J\,=\,1-0$, indicating a molecular structure at $V_{LSR}\,=\,-$61 km/s and possible line broadening
from $-$64 km/s to $-$60 km/s.   They suggested that this structure surrounded Tycho's SNR, and may have been the relic of
a bubble excavated by winds from a single-degenerate progenitor.   However, in their high resolution spectra of Tycho's SNR, Lee et al.
(2007)\cite{Lee07} found the H$\alpha$ emission from the gas ahead of the Balmer-dominated shocks was centered around a completely different radial velocity, $V_{LSR}\,=\,-$35.8$\pm$0.6
km/s.  As mentioned above, Ghavamian et al. (2000) and Lee et al. (2007) determined that this gas was heated by a photoionization precursor, thereby placing it at the same kinematic distance 
as Tycho's SNR and by extension a completely different kinematic distance than the CO clouds observed by Zhou et al. (2016).    This leaves geometric projection as the most likely
explanation for the apparent association between Tycho's SNR and any dense molecular material.   Note that Tian \& Leahy (2011)\cite{TL11} also found no compelling evidence of interaction between 
Tycho and dense molecular cloud material, based on their more recent H~I observations of the SNR.  Although it is clear 
that the eastern side of Tycho is encountering a higher density, more neutral gas than the western side, there is no compelling evidence that the overdense region 
is a molecular cloud, either from H~I observations\cite{Reynoso99} or CO observations.\cite{Zhou16}  

As a final argument, we note that Lee et al. (2007) measured an H$\alpha$ centroid of $V_{LSR}\,=\,-$30.3$\pm$0.2 km/s for the narrow component line in Knot g, 
very similar to that of the photoionization precursor.   The centroid for Knot g measured by 
Ghavamian et al. (2000) was $-$45.6$\pm$1.3 km/s (note the earlier published value of $-$53.9$\pm$1.3 km/s 
was incorrect; see the Erratum to that paper).   Although this value differs from that of Lee et al. (2007), we note that the emission measured
by Ghavamian et al. (2000) was summed over a slit oriented parallel to the Knot g filament, whereas Lee et al. (2007) obtained their spectrum from a slit oriented perpendicular
to the filament, from a localized segment.
Considering the noticeable variations in shock viewing angle along the Knot g filament (e.g., see the HST images)\cite{Lee10} and that the narrow component 
H$\alpha$ line can acquire a bulk Doppler shift up to 10 km/s from the back pressure of cosmic rays immediately ahead of the shock (Wagner et al. 2009),\cite{Wagner09} a cumulative offset
$\sim$15 km/s is plausible between the narrow component centroids by Lee et al. (2007) and Ghavamian et al. (2000).

\section*{Bibliography}
\bibliographystyle{naturemag}

\end{document}